\documentclass[%
superscriptaddress,
twocolumn,
 amsmath,amssymb,
 aps, prl
]{revtex4-2}

\usepackage{graphicx}% Include figure files

\usepackage{dcolumn}% Align table columns on decimal point
\usepackage[dvipsnames]{xcolor}
\usepackage[colorlinks,linkcolor=blue,citecolor=blue,urlcolor=blue]{hyperref}

\usepackage{stix}

\begin{document}

\title{Accurate and efficient simulation of photoemission spectroscopy via Kohn-Sham scattering states}% Force line breaks with

\author{Gian Parusa}
\affiliation{
 PSI Center for Scientific Computing, Theory and Data, Paul Scherrer Institute, 5232 Villigen PSI, Switzerland 
}
\affiliation{
 Department of Physics, University of Fribourg, 1700 Fribourg, Switzerland
}
\affiliation{
 National Centre for Computational Design and Discovery of Novel Materials (MARVEL), Paul Scherrer Institute, 5232 Villigen PSI, Switzerland 
}

\author{Sotirios Fragkos}
\affiliation{
 Université de Bordeaux—CNRS—CEA, CELIA, UMR5107, F33405 Talence, France 
}

\author{Samuel Beaulieu}
\affiliation{
 Université de Bordeaux—CNRS—CEA, CELIA, UMR5107, F33405 Talence, France 
}

\author{Michael Schüler}
\affiliation{
 PSI Center for Scientific Computing, Theory and Data, Paul Scherrer Institute, 5232 Villigen PSI, Switzerland 
}
\affiliation{
 Department of Physics, University of Fribourg, 1700 Fribourg, Switzerland
}
\affiliation{
 National Centre for Computational Design and Discovery of Novel Materials (MARVEL), Paul Scherrer Institute, 5232 Villigen PSI, Switzerland 
}

\email{michael.schueler@psi.ch}

\date{\today}% It is always \today, today,

\begin{abstract}

We introduce an efficient first-principles framework for simulating angle-resolved photoemission spectroscopy (ARPES) based on the direct computation of photoelectron states as solutions of the Kohn-Sham equation with scattering boundary conditions. While the one-step theory of photoemission has a long and successful history, existing implementations are often tied to specialized electronic-structure formalisms. Our approach is formally equivalent to the Lippmann-Schwinger formulation, and it is directly compatible with standard plane-wave and real-space density functional theory codes, enabling seamless integration with advanced exchange-correlation functionals and modern electronic-structure workflows. By providing explicit photoelectron wave functions, the method allows for a transparent analysis of matrix-element effects, multiple scattering, and experimental geometry. We demonstrate the accuracy and predictive power of the framework through circular-dichroism ARPES simulations for monolayer graphene and bulk $2H$-WSe$_2$, achieving excellent agreement with experimental data over a wide photon-energy range. Our results establish a robust and accessible route toward quantitative ARPES modeling, opening the door to systematic studies of orbital textures, many-body effects, and nonequilibrium phenomena within widely used ab initio platforms.
\end{abstract}

\maketitle

%\tableofcontents

% \nocite{*}

\emph{Introduction.---}
ARPES is one of the most powerful experimental techniques to probe the electronic structure of solids~\cite{lv_angle-resolved_2019, yang_visualizing_2018, boschini_time-resolved_2024, sobota_angle-resolved_2021}. The inherent conservation of crystal momentum  in the photoemission process enables ARPES to map the occupied band structure, providing direct insights into the Fermi surface, band topology, surface states, and many-body interactions in materials. Periodic solids are, however, more than just their band structure, as the momentum-dependent orbital character and complex-valued electronic wavefunctions also play a crucial role in determining their physical properties. Paradigmatic examples include the orbital texture in transition metal dichalcogenides~\cite{schusser_towards_2024, heider_geometry-induced_2023, beaulieu_unveiling_2021}, topological insulators~\cite{wang_observation_2011, bentmann_strong_2017, sidilkover_reexamining_2025, wang_circular_2013, bentmann_profiling_2021}, Weyl semimetals~\cite{unzelmann_momentum-space_2021, schusser_assessing_2022, ono_surface_2021, fanciulli_spin_2017, min_orbital_2019} and beyond~\cite{figgemeier_imaging_2025, brinkman_chirality-driven_2024}.

Such wavefunction information is encoded in the ARPES intensity via the photoemission matrix elements, which describe the transition from an initial Bloch state to a final photoelectron state upon interaction with light. It is generally believed that circular dichroism in ARPES (CD-ARPES) is sensitive to the orbital angular momentum and Berry curvature of the initial states~\cite{schuler_local_2020, cho_experimental_2018, cho_studying_2021, takahashi_berry_2015, erhardt_bias-free_2024, kang_measurements_2025}. On the other hand, ARPES is sensitive to details of the final photoelectron states. In particular, the experimental geometry has a profound impact on the ARPES intensity and circular dichroism patterns~\cite{fedchenko_4d_2019}, while various intra-atomic and inter-atomic scattering channels can contribute to the final photoelectron states. The interplay of these contributions leads to complex photon-energy dependent interference effects. Additionally, multiple scattering of the photoelectron wavefunctions in the crystal potential can have a profound impact~\cite{krasovskii_band_2007, strocov_high-energy_2023}. As a result, there is a gap between the intrinsic, i.e., measurement-independent properties of the initial states and the experimentally observed ARPES spectra. 

To bridge this gap, accurate first-principles simulations of ARPES spectra are indispensable. Photoemission theory -- and the one-step model in particular -- has a long and highly successful history~\cite{mahan_theory_1970, feibelman_photoemission_1974, pendry_theory_1976}, providing a rigorous and unified description of excitation, propagation, and escape of the photoelectron within a single quantum-mechanical framework. Among existing implementations, the Korringa-Kohn–Rostoker (KKR) multiple-scattering method is the most mature~\cite{braun_theory_1996, ebert_calculating_2011, minar_calculation_2011}, having been applied to a wide range of materials classes and physical regimes with remarkable success~\cite{durham_theory_1981,arpiainen_circular_2009-1, braun_correlation_2018}. At the same time, the growing complexity of modern ARPES experiments and materials calls for complementary approaches that can be integrated with widely used plane-wave or real-space density functional theory (DFT) codes~\cite{giannozzi_advanced_2017, tancogne-dejean_octopus_2020}, especially in view of advanced electronic structure methods~\cite{onida_electronic_2002,kirchner-hall_extensive_2021, linscott_koopmans_2023}. 
In this spirt, time-dependent density functional theory (TDDFT)~\cite{de_giovannini_ab_2012,de_giovannini_simulating_2013,de_giovannini_first-principles_2017}
has been employed to simulate ARPES spectra; however, its computational cost often limits applications to small systems. The direct calculation of photoelectron states 
in framework of augmented plane waves~\cite{krasovskii_surface_1997,krasovskii_augmented-plane-wave_2004} or high-energy Bloch states~\cite{nozaki_computational_2024} has also been explored, but so far not widely adopted. Ryoo and Park have recently demonstrated how to compute photoelectron states through the Lippmann-Schwinger (LS) equation~\cite{ryoo_lippmann-schwinger_2025}, advancing the compatibility between widely-used DFT implementations and ARPES simulations.

In this work, we present how to compute the final photoelectron states by directly solving the Kohn-Sham equation with appropriate boundary conditions. We show that this approach, which builds on works by Schattke and Krasovskii~\cite{krasovskii_surface_1997, krasovskii_one-step_2022, borghetti_effect_2012, krasovskii_calculation_1999, krasovskii_ab_2021}, is fully equivalent to the LS formalism, directly compatible with standard plane-wave DFT implementations, and significantly more efficient. We also investigate a key aspect of ARPES simulations in the context of plane-wave or real-space DFT, namely the use of pseudopotentials to represent the ionic potential. While the accuracy of pseudopotentials in ground-state calculations has been extensively studied~\cite{prandini_precision_2018}, their accuracy for describing high-energy photoelectron states is less clear. We analyze the impact of different pseudopotentials on the computed ARPES spectra and circular dichroism patterns, demonstrating that high-quality pseudopotentials enable accurate ARPES simulations in the VUV to XUV regime. As concrete examples, we focus on monolayer graphene and bulk $2H$-WSe$_2$ and compare the predicted spectra to published experimental data and new experimental measurements.
% For graphene we compare our results to literature experimental data, while for WSe$_2$ we present new experimental measurements. In both cases, we find excellent qualitative agreement between our simulations and the experimental results, and for graphene even quantitative agreement in the photon-energy dependent CD-ARPES signal. 
% Our work establishes a robust and efficient framework for first-principles ARPES simulations -- to be supported by an upcoming open-source implementation -- that can be widely adopted by the community.

\emph{Theory.---}
Under the sudden approximation, the dipole approximation, and under the assumption of sharp quasiparticle peaks, the ARPES intensity can be expressed as~\cite{hufner_photoelectron_2005},
\begin{align}
    \label{eq:arpes}
    I(\mathbf{k}, E) &= \sum_\alpha f_\alpha(\mathbf{k}) \, \left| \hat{\boldsymbol{\epsilon}} \cdot \boldsymbol{\mathcal{M}}_{\alpha}(\mathbf{k}, E) \right|^2 \delta(\varepsilon_{\alpha}(\mathbf{k}) + \hbar\omega - E) \nonumber \ .
\end{align}
Here, $f_\alpha(\mathbf{k})$ is the occupation function of the Bloch state $|\psi_{\mathbf{k} \alpha} \rangle$ with band index $\alpha$ and crystal momentum $\mathbf{k}$, $\varepsilon_{\alpha}(\mathbf{k})$ is its corresponding energy, $\omega$ is the photon energy, $E$ is the photoelectron energy, $\hat{\boldsymbol{\epsilon}}$ is the polarization vector of light, and $\boldsymbol{\mathcal{M}}_{\alpha}(\mathbf{k}, E)$ is the photoemission matrix element. It is defined as
\begin{align}
    % \label{eq:matrixelement}
    \boldsymbol{\mathcal{M}}_{\alpha}(\mathbf{k}, E) = \langle \chi_{\mathbf{p}} | \mathbf{\hat{\Delta}} | \psi_{\mathbf{k} \alpha} \rangle,
\end{align}
where $\mathbf{\hat{\Delta}}$ is the light-matter coupling operator, and $|\chi_{\mathbf{p}} \rangle$ is the final state with momentum $\mathbf{p}$ and energy $E = \mathbf{p}^2/2$. In-plane momentum conservation implies $\mathbf{p} = \mathbf{k} + p_\perp \, \mathbf{\hat{z}}$ with $\mathbf{\hat{z}}$ being the surface normal direction. 

Both the initial Bloch states and the final photoelectron states can be obtained from the Kohn-Sham (KS) equation $\hat{H}^{\mathrm{KS}} |\psi \rangle = \varepsilon |\psi \rangle$. To clarify, we assume a two-dimensional periodic system~\footnote{The surface of bulk three-dimensional systems can be modeled by slab geometries that are periodic in two dimensions.}. 
Building on previous works~\cite{krasovskii_surface_1997, krasovskii_one-step_2022, borghetti_effect_2012, krasovskii_calculation_1999, krasovskii_ab_2021}, we employ the in-plane Laue representation for wavefunctions:
\begin{align}
    \label{eq:laue}
    \phi_{\mathbf{k}}(\mathbf{r}) = \sum_{\mathbf{G}} e^{i (\mathbf{k} + \mathbf{G}) \cdot \mathbf{r}_{\parallel}} \, f_{\mathbf{k}, \mathbf{G}}(z) .
\end{align}
For the mixed plane-wave/real-space function $f_{\mathbf{k}, \mathbf{G}}(z)$, the KS equation becomes
\begin{align}   
    \label{eq:ks-laue}
    \sum_{\mathbf{G}^\prime} \left[\hat{T}_{\mathbf{G} \mathbf{G}^\prime} + \hat{V}_{\mathbf{G} \mathbf{G}^\prime} \right] f_{\mathbf{k}, \mathbf{G}'}(z) = \varepsilon f_{\mathbf{k}, \mathbf{G}}(z) ,
\end{align}
where 
\begin{align}
    \label{eq:kinetic}
    \hat{T}_{\mathbf{G} \mathbf{G}^\prime} &= \left( \frac{1}{2} (\mathbf{k} + \mathbf{G})^2 - \frac{1}{2} \frac{d^2}{dz^2} \right) \delta_{\mathbf{G} \mathbf{G}^\prime}
\end{align}
is the kinetic energy operator, and $\hat{V}_{\mathbf{G} \mathbf{G}^\prime}$ is the KS potential in Laue representation. 

The key distinction of the final states from the initial states lies in the boundary conditions. Bound states obey $f_{\mathbf{k}, \mathbf{G}}(z) \rightarrow 0$ as $z \rightarrow \pm \infty$, rendering Eq.~\eqref{eq:ks-laue} an eigenvalue problem. In contrast, the final states behave as $f_{\mathbf{k},\mathbf{G}}(z) = e^{i p_\perp z} + r_{\mathbf{k},\mathbf{G}} e^{-i \kappa(\mathbf{G}) z}$ as $z \rightarrow +\infty$, and $f_{\mathbf{k},\mathbf{G}}(z) = t_{\mathbf{k},\mathbf{G}} e^{i \kappa(\mathbf{G}) z}$ as $z \rightarrow -\infty$, where $\kappa(\mathbf{G}) = \sqrt{2 E - (\mathbf{k} + \mathbf{G})^2}$, and $r_{\mathbf{k},\mathbf{G}}$ and $t_{\mathbf{k},\mathbf{G}}$ are unknown coefficients. Only kinematically allowed channels with $2 E > (\mathbf{k} + \mathbf{G})^2$ -- in the language of multi-channel scattering theory~\cite{joachain_quantum_1975}, open channels -- contribute to the asymptotic behavior~\footnote{For semi-infinite systems, $f_{\mathbf{k},\mathbf{G}}(z) \rightarrow 0$ as $z \rightarrow -\infty$.}. Just like in multi-channel scattering theory~\cite{johnson_renormalized_1978}, the asymptotic boundary conditions transform the KS equation~\eqref{eq:ks-laue} into a linear equation for $f_{\mathbf{k},\mathbf{G}}(z)$~\cite{parusa_supplementary_2025}.

Multi-channel scattering theory also establishes a direct link to the LS formalism employed in Ref.~\cite{ryoo_lippmann-schwinger_2025}. Treating the kinetic energy~\eqref{eq:kinetic} as matrix $\hat{\mathbf{T}}$ in $\mathbf{G}$-space, one can readily define the Green's function of the ordinary differential equation $[E\mathbf{I} - \hat{\mathbf{T}}] \mathbf{K}(E; z, z^\prime) = \mathbf{I}\delta(z - z^\prime)$. Since $\hat{\mathbf{T}}$ is diagonal in $\mathbf{G}$-space, the Green's function is also diagonal with elements
\begin{align}
    \label{eq:green}
    K_{\mathbf{G}}(E; z, z^\prime) = -\frac{1}{\alpha_\mathbf{G}(E)} e^{-\alpha_\mathbf{G}(E) |z - z^\prime|} \ .
\end{align}
Here, $\alpha_\mathbf{G}(E) = \sqrt{(\mathbf{k} + \mathbf{G})^2 - 2 E}$ for closed channels and $\alpha_\mathbf{G}(E) = i \kappa(\mathbf{G})$ for open channels. Using this Green's function, one can rewrite Eq.~\eqref{eq:ks-laue} into the LS equation,
\begin{align}
    \label{eq:lippmann-schwinger}
    f_{\mathbf{k}, \mathbf{G}}(z) &= e^{i p_\perp z} \delta_{\mathbf{G}, 0} +\sum_{\mathbf{G}^\prime}\int dz^\prime K_{\mathbf{G}}(E; z, z^\prime) \nonumber \\
    &\quad \times (\hat{V}_{\mathbf{G} \mathbf{G}^\prime}f_{\mathbf{k}, \mathbf{G}^\prime})(z^\prime)  \ .
\end{align}

In practical terms, both the LS equation~\eqref{eq:lippmann-schwinger} and the KS equation~\eqref{eq:ks-laue} with scattering boundary conditions are solved on a discrete $z$-grid with grid spacing $h$, transforming them to linear systems of equations; given a typical problem size, iterative linear solvers are employed. We have implemented both approaches with higher-order discretization up to $\mathcal{O}(h^5)$ and verified their numerical equivalence~\cite{parusa_supplementary_2025}. However, the route via the KS equation offers the advantage of highly effective preconditioning strategies akin to those used in iterative diagonalization, leading to significantly faster convergence.

% The reformulation in terms of a linear system of equations also allows us to directly include the non-local part of the pseudopotential, as well as an optical potential to account for inelastic effects.

\emph{Results.---}
To demonstrate the capabilities of our method, we present theoretical ARPES results obtained by solving the KS equations for monolayer graphene and bulk 2H-WSe$_2$ and compare them with published~\cite{gierz_graphene_2012} and new experimental data. In particular, we focus on dichroism as the key observable. In particular, the circular dirchroism in the angular distribution (CDAD) is highly sensitive to the detailed structure of the final states, making it a stringent and well-suited benchmark for assessing the accuracy of our approach.

\begin{figure}[t]
\includegraphics[width=\columnwidth]{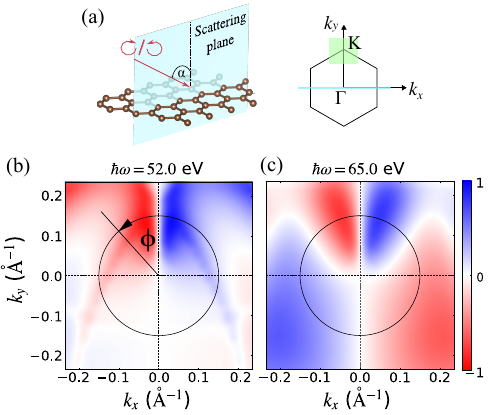}% Here is how to import EPS art
\caption{\label{fig:1} (a) Experimental geometry used in Ref.~\cite{gierz_graphene_2012}. The angle of incidence is $\alpha=50^\circ$. 
(b) Sketch of the Brillouin zone of graphene with the region of interest highlighted in green around the $K$-point.
(c) Calculated energy-integrated normalized CDAD around the $K$-point of graphene at photon energy 52 eV, and (d) at photon energy 65 eV. Results are obtained with all-electron method. }
\end{figure}

\begin{figure}[t]
\includegraphics{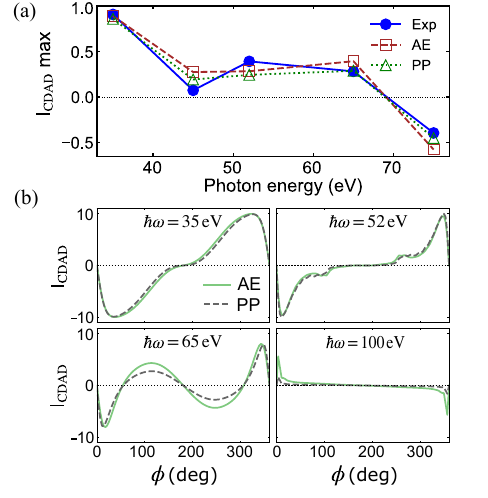}
\caption{\label{fig:2} (a) Comparison of the maximum intensity of the normalized CDAD at the Fermi level along $k_y = 0$ from the experiments, all-electron (AE) and pseudopotential (PP) methods. The experimental data are extracted from Ref.~\cite{gierz_graphene_2012}, with minus sign indicating CDAD sign reversal. (b) Angular dependence of the normalized CDAD spectra for different photon energies. The angle $\phi$ is defined in Fig.~\ref{fig:1}(b).}
\end{figure}

For graphene, we compare our theoretical method against the experimental results from Ref.~\cite{gierz_graphene_2012} and adopt the same experimental geometry as illustrated in Fig.~\ref{fig:1}(a). The calculations are performed using both all-electron (AE) and pseudopotential (PP) methods~\cite{parusa_supplementary_2025}. As in Ref.~\cite{gierz_graphene_2012}, we focus on the CDAD close to the $K$-point of graphene (see Fig.~\ref{fig:1}(b)), and we present results for the energy-integrated CDAD normalized to the total intensity. 
Fig.~\ref{fig:1}(c) and (d) show the results for photon energies 52 eV and 65 eV, respectively.
%
% Time-reversal symmetry dictates that CDAD must be antisymmetric $I_{\mathrm{CDAD}} (\mathbf{k}) = - I_{\mathrm{CDAD}} (-\mathbf{k})$, and the results show this feature perfectly. 
%
In Fig.~\ref{fig:1}(c), we successfully reproduce the nodal line (where the CDAD spectrum vanishes) along the $\Gamma-K$ direction as well as the overall CDAD texture. There are, however, more pronounced "leg" features in the calculation as compared to the experiment. In Fig.~\ref{fig:1}(d), we also reproduce the $\Gamma-K$ nodal line as well as the other two nodal lines on the first and second quadrants. The overall texture is in remarkable agreement with the experiment. Results for additional photon energies are presented in the Supplementary Material~\cite{parusa_supplementary_2025}.
We also analyzed the maximum intensity of the normalized CDAD as a function of photon energy, following the same convention as Fig.~2(b) of Ref.~\cite{gierz_graphene_2012}. Our results are shown in Fig.~\ref{fig:2}(a) together with the experimental data. We successfully reproduce the reduction of the maximum intensity at 45 eV as well as the CDAD sign reversal at 75 eV, improving significantly over the KKR results also presented in Ref.~\cite{gierz_graphene_2012}.

So far, we have used the AE method to compute the photoelectron states. For practical applications, however, PP methods are more widely used due to their efficiency. We now analyze the accuracy of PP methods for simulating ARPES spectra by comparing them to the AE results. We use the generalized norm-conserving PP from the \textsc{PseudoDojo} library~\cite{setten_pseudodojo_2018}. The photon-energy-dependent CDAD maximum intensity from PP is also shown in Fig.~\ref{fig:2}(a). We observe an overall excellent agreement with the AE results over the whole photon energy range. In Fig.~\ref{fig:2}(b), we compare the methods in more detail by analyzing the normalized CDAD along a circle of radius $r = 0.15$ \AA$^{-1}$ around the $K$-point, as illustrated in Fig.~\ref{fig:1}(c). The results for PP agree well with those of AE up to photon energy 100 eV; small deviations start to appear at 65 eV. Nevertheless, the PP results successfully reproduce all the AE features, including the nodal line positions. The success of the PP method can be attributed to its non-local part, which by construction is optimized to reproduce the atomic AE scattering properties~\cite{vanderbilt_soft_1990}.

The non-local part of the PP is indeed crucial for reproducing the scattering properties of the photoelectron states. In general, the observed spectra change drastically when one switches off this contribution in the photoelectron state calculation~\cite{parusa_supplementary_2025}. However, for a photon energy of 52 eV, the difference between including and excluding the non-local part is negligible. This is because the final states are predominantly composed of $d$-like partial waves around the Carbon atoms, which are not affected by the non-local part of the PP that only acts on $s$ and $p$ angular momentum channels~\cite{schlipf_optimization_2015}. 
%
% possess a $d$-like partial wave character rather than $s$ or $p$, which has already been noted in Ref.~\cite{Gierz2}. Since the non-local part of the Carbon PP only fixes the $s$ and $p$ angular momentum channel, $\hat{V}_{\mathrm{NL}} |\chi_E \rangle \simeq 0$ around  $\hbar \omega =  52$ eV and hence only the local part of the PP contributes, which in general is optimized for angular momentum channel more than $\ell_{\mathrm{max}}$ (with $\ell_{\mathrm{max}}$ being the maximum angular momentum quantum number in $V_{\mathrm{NL}}$)~\cite{Schlipf-loc}.
%
In contrast, non-local contributions to the light-matter coupling operator $\mathbf{\hat{\Delta}}$~\footnote{We use the velocity operator $\mathbf{\hat{\Delta}} = i [\hat{H}^\mathrm{KS}, \hat{\mathbf{r}}]$. } are negligible in all cases~\cite{parusa_supplementary_2025}.

\begin{figure}[t]
\includegraphics{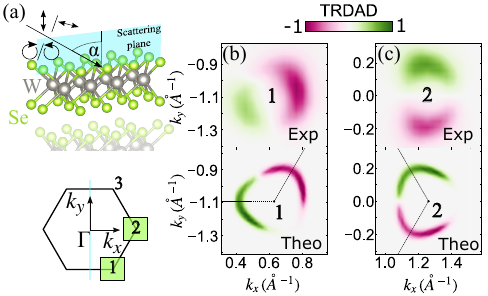}% Here is how to import EPS art
\caption{\label{fig:3} (a) Experimental geometry for WSe$_2$. In real space (top panel), the scattering plane is in indicated by the vertical plane while in the reciprocal space (bottom panel) it is indicated by the vertical line along the the $k_y$ direction. The angle of incidence is $\alpha = 65^{\circ}$. (b)--(e) show the TRDAD spectra of the experiment and the theory with $E - E_{\mathrm{VBM}} = -0.25$ eV in the vicinity of K-points 1 and 2, respectively.}
\end{figure}

As a second example, we present new experimental and theoretical results for bulk 2H-WSe$_2$. The 2H-WSe$_2$ experimental data have been acquired using the time- and polarization-resolved extreme ultraviolet momentum microscope instrument~\cite{fragkos_time-_2025} at the Centre Lasers Intenses et Applications, in Bordeaux, France. Details about the experimental setup can be found elsewhere~\cite{fragkos_time-_2025} and in Ref.~\cite{parusa_supplementary_2025}. The experimental geometry is shown in Fig.~\ref{fig:3}(a).
%
% depending on if we want to show more data I can modify this part above. 
%
% The experimental setup is presented in Ref.~\cite{parusa_supplementary_2025}. 
All calculations are performed with PP method. To highlight the importance of the non-local part of the PP, we compare two variants of PPs: (i) including the semicore states ($5s$, $5p$) for W and $3d$ for Se, and (ii) excluding them.

As the first test, we compute the ARPES spectra with $p$-polarized light and analyze the dark corridor at binding energies close to the valence band maximum. Our calculations~\cite{parusa_supplementary_2025} with both PP variants reproduce the experimental spectra well, including the rotation of the dark corridor upon crystal rotation by $60^\circ$. Taking the difference of the normalized spectra at $0^\circ$ and $60^\circ$ crystal rotation defines the time-reversal dichroism in angular distribution (TRDAD)~\cite{beaulieu_revealing_2020}. The TRDAD obtained from experiments (Fig.~\ref{fig:3}(b), (c)) is in excellent agreement with our simulations (Fig.~\ref{fig:3}(d), (e)).

Next we inspect the CDAD, shown in Fig.~\ref{fig:3}(a) at binding energy $E - E_{\mathrm{VBM}} = -0.25$ eV. We note that PP method with semicore states successfully reproduces the CDAD texture at $K$-point 2 and 3. This is already different for the PP without semicore states, which only agrees at $K$-point 3. 
% Nevertheless, both PP variants fail to fully reproduce the CDAD texture at $K$-point 1, albeit the results with semicore states are closer to the experiment. 

For a quantitative comparison, we analyze the valley-integrated CDAD (valley CDAD) at each $K$-point, which is generally a more robust observable than the full angular distribution. The valley CDAD, presented in Fig.~\ref{fig:3}(g)--(i) as function of the binding energy, obtained with PP method with semicorestates is in almost quantitative agreement with the experiment for all three $K$-points (note that the experimental and calculated data are normalized the same way). In contrast, the valley CDAD from PP without semicore states deviates from the experiment, especially at $K$-point 1 and 2, where the valley CDAD of $K$-point 2 appears to be significantly lower than that of the experiment and thus appears below that of $K$-point 1. This analysis confirms that including semicore states in the PP is essential for accurately describing the scattering phase of the photoelectron states around each atomic site, which in turn determines the angular distribution through the interferometric nature of the photoemission process~\cite{moser_toy_2023, kern_simple_2023, yen_controllable_2024, boban_scattering_2025, sidilkover_reexamining_2025, heider_geometry-induced_2023}.

\begin{figure}[t]
\includegraphics{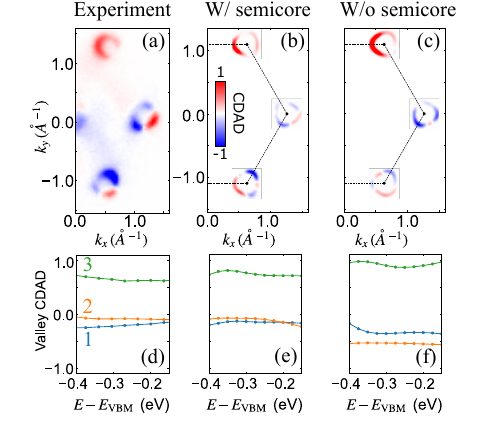}% Here is how to import EPS art
\caption{\label{fig:4} (a)--(c) CDAD texture at binding energy $E - E_{\mathrm{VBM}} = -0.25$ eV. The spectrum is antisymmetric with respect to the scattering plane and thus only half of the Brillouin zone is shown.  (d) -- (f) valley-integrated CDAD as a function of binding energy for the experiment, PP with and without semicore states, respectively.}
\end{figure}

% We then perform an integration over the angular distribution of the circular dichroism signal at each $K$-point and observe their modulation against the binding energy, as plotted in Fig.~\ref{fig:3}(b). The $x$-axis is shifted so that the zero aligns with the binding energy of the valence band maximum (VBM). There is a consistently flat behavior over all $K$-points as shown by the experiment. Interestingly, even though the semicore PP predicts a wrong texture at $K$-point 1, its valley-integrated CDAD resembles closely with that of the experiment. Their numerical values agree pretty well over the energy window even though a slight crossover between $K$-point 1 and 2 happens around $E-E_{\mathrm{VBM}} = -0.2$. On the other hand, the non-semicore PP shows valley CDAD’s that are quite off from the experiment for $K$-point 1 and 2, and exhibits more modulation at $K$-point 3. These analyses imply that semicore exclusion affects the photoelectron states significantly due to less accurate description of the scattering properties. Semicore inclusion improves the number of angular momentum channels in the scattered wave and hence achieves better scattering phase agreement.

\emph{Discussion.---}
In summary, we have presented a method to compute the final photoelectron states by solving the Kohn-Sham equation with appropriate scattering boundary conditions. This approach is fully equivalent to the Lippmann-Schwinger formalism, but significantly more efficient due to the possibility of employing effective preconditioning strategies from iterative diagonalization. We demonstrated that the PP method can accurately describe photoelectron states, provided the pseudoatomic scattering phase matches the AE one. Recent developments of high-accuracy PPs~\cite{bosoni_how_2024} are encouraging in this regard.

In contrast to other methods that yield the photoemission intensity only, calculating the photoelectron states $|\chi_{\mathbf{p}} \rangle$ from the Kohn-Sham equation explicitly offers a number of advantages: (i) the role of multiple scattering and interference effects in the final states can be directly analyzed, (ii) the matrix elements can be computed for arbitrary experimental geometries and light polarizations as post-processing, and (iii) the intrinsic properties of the initial states can be disentangled from extrinsic effects related to the final states. 
To treat correlation effects beyond DFT, the presented method can be combined with Green's function techniques. For instance, the Green's function $G^<_{\alpha \alpha^\prime}(\mathbf{k},\omega)$ obtained from a $GW$ calculation~\cite{onida_electronic_2002} defines the ARPES intensity via $I(\mathbf{k},E) \propto \sum_{\alpha\alpha^\prime}\mathcal{M}_\alpha(\mathbf{k},E)G^<_{\alpha \alpha^\prime}(\mathbf{k},\omega-E) \mathcal{M}^*_{\alpha^\prime}(\mathbf{k},E)$. 
The presented methodology comes with a few downsides as well. On the technical side, treating surfaces with large unit cells may be out of reach computationally. In particular, truly semi-infinite systems would require would a Green's function formulation such as the KKR method. At higher photon energies, the pseudopotential method may also become less accurate due to the increasing importance of core states. In such cases, the reconstruction of the all-electron wavefunctions from the pseudo-wavefunctions~\cite{kresse_ultrasoft_1999} could be a viable solution.

Finally, \emph{ab initio} description of pump-probe time-resolved ARPES (trARPES) is also possible~\cite{kern_photoemission_2023}. KKR-based implementations of trARPES have been developed~\cite{braun_one-step_2015}, but they are limited to standard DFT. The non-equilibrium Green's function formalism~\cite{stefanucci_nonequilibrium_2013}, combining many-body perturbation theory and ultrafast dynamics, is the most promise route towards predictive simulations of trARPES~\cite{freericks_theoretical_2009, schuler_theory_2021}. The photoemission matrix elements are essential ingredients in this framework.
We have demonstrated this approach for Floquet states in light-driven graphene in Ref.~\cite{parusa_supplementary_2025} under pump-probe conditions as in recent experiments~\cite{choi_observation_2025, merboldt_observation_2025}.

The direct calculation of photoelectron states and matrix elements in an efficient and accessible framework is thus a major step towards the interpretation of ARPES and understanding of materials in and out of equilibrium.

% The results with AE or PP produce the correct symmetry and excellently give the final states description that are close to that of the experiment. As demonstrated by the results for graphene, semi-local PP whose non-local part carries accurate scattering properties can reproduce the AE spectra up to soft X-ray regime. The scattering information encoded in the non-local part is proven to be essential for this matter, and our simulation with WSe$_2$ confirms that fixing more angular momentum channel increases the accuracy of the resulting final states.

\begin{acknowledgments}

\emph{Acknowledgments.---}
This research was supported by the NCCR MARVEL, a National Centre of Competence in Research, funded by the Swiss National Science Foundation (grant number 205602).

\end{acknowledgments}

\bibliographystyle{apsrev4-2}
\bibliography{zotero}% Produces the bibliography via BibTeX.

@book{stefanucci_nonequilibrium_2013,
   title       = {Nonequilibrium {Many}-{Body} {Theory} of {Quantum} {Systems}: {A} {Modern} {Introduction}},
   publisher   = {Cambridge University Press},
   author      = {Stefanucci, Gianluca and Leeuwen, Robert van},
   year        = {2013},
}

@article{kresse_ultrasoft_1999,
  title = {From ultrasoft pseudopotentials to the projector augmented-wave method},
  author = {Kresse, G. and Joubert, D.},
  journal = {Phys. Rev. B},
  volume = {59},
  issue = {3},
  pages = {1758--1775},
  numpages = {0},
  year = {1999},
  month = {Jan},
  publisher = {American Physical Society},
  doi = {10.1103/PhysRevB.59.1758},
  url = {https://link.aps.org/doi/10.1103/PhysRevB.59.1758}
}

@article{prandini_precision_2018,
   title       = {Precision and efficiency in solid-state pseudopotential calculations},
   volume      = {4},
   url         = {https://www.nature.com/articles/s41524-018-0127-2},
   doi         = {10.1038/s41524-018-0127-2},
   number      = {1},
   journal     = {npj Comput Mater},
   author      = {Prandini, Gianluca and Marrazzo, Antimo and Castelli, Ivano E. and Mounet, Nicolas and Marzari, Nicola},
   year        = {2018},
   pages       = {72},
}

@article{linscott_koopmans_2023,
   title       = {koopmans : {An} {Open}-{Source} {Package} for {Accurately} and {Efficiently} {Predicting} {Spectral} {Properties} with {Koopmans} {Functionals}},
   volume      = {19},
   copyright   = {https://creativecommons.org/licenses/by/4.0/},
   url         = {https://pubs.acs.org/doi/10.1021/acs.jctc.3c00652},
   doi         = {10.1021/acs.jctc.3c00652},
   number      = {20},
   journal     = {J. Chem. Theor. Comput.},
   author      = {Linscott, Edward B. and Colonna, Nicola and De Gennaro, Riccardo and Nguyen, Ngoc Linh and Borghi, Giovanni and Ferretti, Andrea and Dabo, Ismaila and Marzari, Nicola},
   year        = {2023},
   pages       = {7097--7111},
}

@article{kirchner-hall_extensive_2021,
   title       = {Extensive {Benchmarking} of {DFT}+{U} {Calculations} for {Predicting} {Band} {Gaps}},
   volume      = {11},
   url         = {https://www.mdpi.com/2076-3417/11/5/2395},
   doi         = {10.3390/app11052395},
   number      = {5},
   journal     = {Applied Sciences},
   author      = {Kirchner-Hall, Nicole E. and Zhao, Wayne and Xiong, Yihuang and Timrov, Iurii and Dabo, Ismaila},
   year        = {2021},
   pages       = {2395},
}

@article{krasovskii_augmented-plane-wave_2004,
   title       = {Augmented-plane-wave approach to scattering of {Bloch} electrons by an interface},
   volume      = {70},
   copyright   = {http://link.aps.org/licenses/aps-default-license},
   url         = {https://link.aps.org/doi/10.1103/PhysRevB.70.245322},
   doi         = {10.1103/PhysRevB.70.245322},
   number      = {24},
   journal     = {Phys. Rev. B},
   author      = {Krasovskii, E. E.},
   year        = {2004},
   pages       = {245322},
}

@article{arpiainen_circular_2009-1,
   title       = {Circular {Dichroism} in the {Angle}-{Resolved} {Photoemission} {Spectrum} of the {High}-{Temperature} {Bi} 2 {Sr} 2 {CaCu} 2 {O} 8 + $\delta$ {Superconductor}: {Can} {These} {Measurements} {Be} {Interpreted} as {Evidence} for {Time}-{Reversal} {Symmetry} {Breaking}?},
   volume      = {103},
   copyright   = {http://link.aps.org/licenses/aps-default-license},
   url         = {https://link.aps.org/doi/10.1103/PhysRevLett.103.067005},
   doi         = {10.1103/PhysRevLett.103.067005},
   number      = {6},
   journal     = {Phys. Rev. Lett.},
   author      = {Arpiainen, V. and Bansil, A. and Lindroos, M.},
   year        = {2009},
   pages       = {067005},
}

@article{durham_theory_1981,
   title       = {Theory of photoemission from random alloys},
   volume      = {11},
   url         = {https://iopscience.iop.org/article/10.1088/0305-4608/11/11/027},
   doi         = {10.1088/0305-4608/11/11/027},
   number      = {11},
   journal     = {J. Phys. F: Met. Phys.},
   author      = {Durham, P J},
   year        = {1981},
   pages       = {2475--2493},
}

@article{braun_one-step_2015,
   title       = {One-step theory of pump-probe photoemission},
   volume      = {91},
   copyright   = {http://link.aps.org/licenses/aps-default-license},
   url         = {https://link.aps.org/doi/10.1103/PhysRevB.91.035119},
   doi         = {10.1103/PhysRevB.91.035119},
   number      = {3},
   journal     = {Phys. Rev. B},
   author      = {Braun, J. and Rausch, R. and Potthoff, M. and Min\'{a}r, J. and Ebert, H.},
   year        = {2015},
   pages       = {035119},
}

@article{braun_theory_1996,
   title       = {The theory of angle-resolved ultraviolet photoemission and its applications to ordered materials},
   volume      = {59},
   url         = {https://iopscience.iop.org/article/10.1088/0034-4885/59/10/002},
   doi         = {10.1088/0034-4885/59/10/002},
   number      = {10},
   journal     = {Rep. Prog. Phys.},
   author      = {Braun, J},
   year        = {1996},
   pages       = {1267--1338},
}

@article{braun_correlation_2018,
   title       = {Correlation, temperature and disorder: {Recent} developments in the one-step description of angle-resolved photoemission},
   volume      = {740},
   url         = {https://www.sciencedirect.com/science/article/pii/S0370157318300358},
   doi         = {10.1016/j.physrep.2018.02.007},
   journal     = {Phys. Rep.},
   author      = {Braun, J\"{u}rgen and Min\'{a}r, J\'{a}n and Ebert, Hubert},
   year        = {2018},
   pages       = {1--34},
}

@article{fragkos_time-_2025,
	title = {Time- and polarization-resolved extreme ultraviolet momentum microscopy},
	volume = {96},
	issn = {0034-6748},
	url = {https://doi.org/10.1063/5.0260193},
	doi = {10.1063/5.0260193},
	number = {11},
	urldate = {2025-12-19},
	journal = {Review of Scientific Instruments},
	author = {Fragkos, Sotirios and Courtade, Quentin and Tkach, Olena and Gaudin, Jérôme and Descamps, Dominique and Barrette, Guillaume and Petit, Stéphane and Schönhense, Gerd and Mairesse, Yann and Beaulieu, Samuel},
	month = nov,
	year = {2025},
	pages = {115201},
}

@book{joachain_quantum_1975,
	address = {Amsterdam, The Netherlands},
	title = {Quantum collision theory},
	isbn = {0-7204-0294-8},
	url = {https://inis.iaea.org/records/xzec9-69q24},
	urldate = {2025-12-15},
	publisher = {North-Holland},
	author = {Joachain, C. J.},
	year = {1975},
}

@article{johnson_renormalized_1978,
	title = {The renormalized {Numerov} method applied to calculating bound states of the coupled‐channel {Schroedinger} equation},
	volume = {69},
	issn = {0021-9606},
	url = {https://doi.org/10.1063/1.436421},
	doi = {10.1063/1.436421},
	number = {10},
	urldate = {2025-12-18},
	journal = {The Journal of Chemical Physics},
	author = {Johnson, B. R.},
	month = nov,
	year = {1978},
	pages = {4678--4688},
}

@article{beaulieu_unveiling_2021,
	title = {Unveiling the orbital texture of {1T}-{TiTe2} using intrinsic linear dichroism in multidimensional photoemission spectroscopy},
	volume = {6},
	copyright = {2021 The Author(s)},
	issn = {2397-4648},
	url = {https://www.nature.com/articles/s41535-021-00398-3},
	doi = {10.1038/s41535-021-00398-3},
	language = {en},
	number = {1},
	urldate = {2025-12-17},
	journal = {npj Quantum Materials},
	author = {Beaulieu, Samuel and Schüler, Michael and Schusser, Jakub and Dong, Shuo and Pincelli, Tommaso and Maklar, Julian and Neef, Alexander and Reinert, Friedrich and Wolf, Martin and Rettig, Laurenz and Minár, Ján and Ernstorfer, Ralph},
	month = nov,
	year = {2021},
	pages = {93},
}

@article{boschini_time-resolved_2024,
	title = {Time-resolved {ARPES} studies of quantum materials},
	volume = {96},
	url = {https://link.aps.org/doi/10.1103/RevModPhys.96.015003},
	doi = {10.1103/RevModPhys.96.015003},
	number = {1},
	journal = {Rev. Mod. Phys.},
	author = {Boschini, Fabio and Zonno, Marta and Damascelli, Andrea},
	month = feb,
	year = {2024},
	pages = {015003},
}

@article{mahan_theory_1970,
	title = {Theory of {Photoemission} in {Simple} {Metals}},
	volume = {2},
	url = {https://link.aps.org/doi/10.1103/PhysRevB.2.4334},
	doi = {10.1103/PhysRevB.2.4334},
	number = {11},
	journal = {Phys. Rev. B},
	author = {Mahan, G. D.},
	month = dec,
	year = {1970},
	pages = {4334--4350},
}

@article{freericks_theoretical_2009,
	title = {Theoretical {Description} of {Time}-{Resolved} {Photoemission} {Spectroscopy}: {Application} to {Pump}-{Probe} {Experiments}},
	volume = {102},
	shorttitle = {Theoretical {Description} of {Time}-{Resolved} {Photoemission} {Spectroscopy}},
	url = {https://link.aps.org/doi/10.1103/PhysRevLett.102.136401},
	doi = {10.1103/PhysRevLett.102.136401},
	number = {13},
	urldate = {2025-12-17},
	journal = {Physical Review Letters},
	author = {Freericks, J. K. and Krishnamurthy, H. R. and Pruschke, Th.},
	month = mar,
	year = {2009},
	pages = {136401},
}

@article{krasovskii_surface_1997,
	title = {Surface electronic structure with the linear methods of band theory},
	volume = {56},
	url = {https://link.aps.org/doi/10.1103/PhysRevB.56.12874},
	doi = {10.1103/PhysRevB.56.12874},
	number = {20},
	urldate = {2025-12-17},
	journal = {Physical Review B},
	author = {Krasovskii, E. E. and Schattke, W.},
	month = nov,
	year = {1997},
	pages = {12874--12883},
}

@article{ono_surface_2021,
	title = {Surface band characters of the {Weyl} semimetal candidate material \$\{{\textbackslash}mathrm\{{MoTe}\}\}\_\{2\}\$ revealed by one-step angle-resolved photoemission theory},
	volume = {103},
	url = {https://link.aps.org/doi/10.1103/PhysRevB.103.125139},
	doi = {10.1103/PhysRevB.103.125139},
	number = {12},
	urldate = {2025-12-17},
	journal = {Physical Review B},
	author = {Ono, Ryota and Marmodoro, Alberto and Schusser, Jakub and Nakata, Yoshitaka and Schwier, Eike F. and Braun, Jürgen and Ebert, Hubert and Minár, Ján and Sakamoto, Kazuyuki and Krüger, Peter},
	month = mar,
	year = {2021},
	pages = {125139},
}

@article{bentmann_strong_2017,
	title = {Strong {Linear} {Dichroism} in {Spin}-{Polarized} {Photoemission} from {Spin}-{Orbit}-{Coupled} {Surface} {States}},
	volume = {119},
	url = {https://link.aps.org/doi/10.1103/PhysRevLett.119.106401},
	doi = {10.1103/PhysRevLett.119.106401},
	number = {10},
	urldate = {2025-12-17},
	journal = {Physical Review Letters},
	author = {Bentmann, H. and Maaß, H. and Krasovskii, E. E. and Peixoto, T. R. F. and Seibel, C. and Leandersson, M. and Balasubramanian, T. and Reinert, F.},
	month = sep,
	year = {2017},
	pages = {106401},
}

@article{fanciulli_spin_2017,
	title = {Spin {Polarization} and {Attosecond} {Time} {Delay} in {Photoemission} from {Spin} {Degenerate} {States} of {Solids}},
	volume = {118},
	url = {https://link.aps.org/doi/10.1103/PhysRevLett.118.067402},
	doi = {10.1103/PhysRevLett.118.067402},
	number = {6},
	urldate = {2025-12-17},
	journal = {Physical Review Letters},
	author = {Fanciulli, Mauro and Volfová, Henrieta and Muff, Stefan and Braun, Jürgen and Ebert, Hubert and Minár, Jan and Heinzmann, Ulrich and Dil, J. Hugo},
	month = feb,
	year = {2017},
	pages = {067402},
}

@article{de_giovannini_simulating_2013,
	title = {Simulating {Pump}–{Probe} {Photoelectron} and {Absorption} {Spectroscopy} on the {Attosecond} {Timescale} with {Time}-{Dependent} {Density} {Functional} {Theory}},
	volume = {14},
	copyright = {Copyright © 2013 WILEY-VCH Verlag GmbH \& Co. KGaA, Weinheim},
	issn = {1439-7641},
	url = {https://onlinelibrary.wiley.com/doi/abs/10.1002/cphc.201201007},
	doi = {10.1002/cphc.201201007},
	language = {en},
	number = {7},
	urldate = {2025-12-17},
	journal = {ChemPhysChem},
	author = {De Giovannini, Umberto and Brunetto, Gustavo and Castro, Alberto and Walkenhorst, Jessica and Rubio, Angel},
	year = {2013},
	pages = {1363--1376},
}

@article{kern_simple_2023,
	title = {Simple extension of the plane-wave final state in photoemission: {Bringing} understanding to the photon-energy dependence of two-dimensional materials},
	volume = {5},
	shorttitle = {Simple extension of the plane-wave final state in photoemission},
	url = {https://link.aps.org/doi/10.1103/PhysRevResearch.5.033075},
	doi = {10.1103/PhysRevResearch.5.033075},
	number = {3},
	urldate = {2025-12-17},
	journal = {Physical Review Research},
	author = {Kern, Christian S. and Haags, Anja and Egger, Larissa and Yang, Xiaosheng and Kirschner, Hans and Wolff, Susanne and Seyller, Thomas and Gottwald, Alexander and Richter, Mathias and De Giovannini, Umberto and Rubio, Angel and Ramsey, Michael G. and Bocquet, François C. and Soubatch, Serguei and Tautz, F. Stefan and Puschnig, Peter and Moser, Simon},
	month = aug,
	year = {2023},
	pages = {033075},
}

@article{boban_scattering_2025,
	title = {Scattering makes a difference in circular dichroic angle-resolved photoemission},
	volume = {111},
	url = {https://link.aps.org/doi/10.1103/PhysRevB.111.115127},
	doi = {10.1103/PhysRevB.111.115127},
	number = {11},
	urldate = {2025-12-17},
	journal = {Physical Review B},
	author = {Boban, Honey and Qahosh, Mohammed and Hou, Xiao and Sobol, Tomasz and Beyer, Edyta and Szczepanik, Magdalena and Baranowski, Daniel and Mearini, Simone and Feyer, Vitaliy and Mokrousov, Yuriy and Zhou, Yishui and Su, Yixi and Jin, Keda and Wichmann, Tobias and Martinez-Castro, Jose and Ternes, Markus and Tautz, F. Stefan and Lüpke, Felix and Schneider, Claus M. and Henk, Jürgen and Plucinski, Lukasz},
	month = mar,
	year = {2025},
	pages = {115127},
}

@article{beaulieu_revealing_2020,
	title = {Revealing {Hidden} {Orbital} {Pseudospin} {Texture} with {Time}-{Reversal} {Dichroism} in {Photoelectron} {Angular} {Distributions}},
	volume = {125},
	url = {https://link.aps.org/doi/10.1103/PhysRevLett.125.216404},
	doi = {10.1103/PhysRevLett.125.216404},
	number = {21},
	journal = {Phys. Rev. Lett.},
	author = {Beaulieu, S. and Schusser, J. and Dong, S. and Schüler, M. and Pincelli, T. and Dendzik, M. and Maklar, J. and Neef, A. and Ebert, H. and Hricovini, K. and Wolf, M. and Braun, J. and Rettig, L. and Minár, J. and Ernstorfer, R.},
	month = nov,
	year = {2020},
	pages = {216404},
}

@article{bentmann_profiling_2021,
	title = {Profiling spin and orbital texture of a topological insulator in full momentum space},
	volume = {103},
	url = {https://link.aps.org/doi/10.1103/PhysRevB.103.L161107},
	doi = {10.1103/PhysRevB.103.L161107},
	number = {16},
	urldate = {2025-12-17},
	journal = {Physical Review B},
	author = {Bentmann, H. and Maaß, H. and Braun, J. and Seibel, C. and Kokh, K. A. and Tereshchenko, O. E. and Schreyeck, S. and Brunner, K. and Molenkamp, L. W. and Miyamoto, K. and Arita, M. and Shimada, K. and Okuda, T. and Kirschner, J. and Tusche, C. and Ebert, H. and Minár, J. and Reinert, F.},
	month = apr,
	year = {2021},
	pages = {L161107},
}

@article{feibelman_photoemission_1974,
	title = {Photoemission spectroscopy—{Correspondence} between quantum theory and experimental phenomenology},
	volume = {10},
	url = {https://link.aps.org/doi/10.1103/PhysRevB.10.4932},
	doi = {10.1103/PhysRevB.10.4932},
	number = {12},
	journal = {Phys. Rev. B},
	author = {Feibelman, Peter J. and Eastman, D. E.},
	month = dec,
	year = {1974},
	pages = {4932--4947},
}

@article{kern_photoemission_2023,
	title = {Photoemission orbital tomography for excitons in organic molecules},
	volume = {108},
	url = {https://link.aps.org/doi/10.1103/PhysRevB.108.085132},
	doi = {10.1103/PhysRevB.108.085132},
	number = {8},
	urldate = {2025-12-17},
	journal = {Physical Review B},
	author = {Kern, C. S. and Windischbacher, A. and Puschnig, P.},
	month = aug,
	year = {2023},
	pages = {085132},
}

@article{min_orbital_2019,
	title = {Orbital {Fingerprint} of {Topological} {Fermi} {Arcs} in the {Weyl} {Semimetal} {TaP}},
	volume = {122},
	url = {https://link.aps.org/doi/10.1103/PhysRevLett.122.116402},
	doi = {10.1103/PhysRevLett.122.116402},
	number = {11},
	urldate = {2025-12-17},
	journal = {Physical Review Letters},
	author = {Min, Chul-Hee and Bentmann, Hendrik and Neu, Jennifer N. and Eck, Philipp and Moser, Simon and Figgemeier, Tim and Ünzelmann, Maximilian and Kissner, Katharina and Lutz, Peter and Koch, Roland J. and Jozwiak, Chris and Bostwick, Aaron and Rotenberg, Eli and Thomale, Ronny and Sangiovanni, Giorgio and Siegrist, Theo and Di Sante, Domenico and Reinert, Friedrich},
	month = mar,
	year = {2019},
	pages = {116402},
}

@article{krasovskii_one-step_2022,
	title = {One-{Step} {Theory} {View} on {Photoelectron} {Diffraction}: {Application} to {Graphene}},
	volume = {12},
	copyright = {http://creativecommons.org/licenses/by/3.0/},
	issn = {2079-4991},
	shorttitle = {One-{Step} {Theory} {View} on {Photoelectron} {Diffraction}},
	url = {https://www.mdpi.com/2079-4991/12/22/4040},
	doi = {10.3390/nano12224040},
	language = {en},
	number = {22},
	urldate = {2025-12-17},
	journal = {Nanomaterials},
	author = {Krasovskii, Eugene},
	month = nov,
	year = {2022},
}

@article{choi_observation_2025,
	title = {Observation of {Floquet}–{Bloch} states in monolayer graphene},
	volume = {21},
	copyright = {2025 The Author(s), under exclusive licence to Springer Nature Limited},
	issn = {1745-2481},
	url = {https://www.nature.com/articles/s41567-025-02888-8},
	doi = {10.1038/s41567-025-02888-8},
	language = {en},
	number = {7},
	urldate = {2025-12-17},
	journal = {Nature Physics},
	author = {Choi, Dongsung and Mogi, Masataka and De Giovannini, Umberto and Azoury, Doron and Lv, Baiqing and Su, Yifan and Hübener, Hannes and Rubio, Angel and Gedik, Nuh},
	month = jul,
	year = {2025},
	pages = {1100--1105},
}

@article{merboldt_observation_2025,
	title = {Observation of {Floquet} states in graphene},
	volume = {21},
	copyright = {2025 The Author(s)},
	issn = {1745-2481},
	url = {https://www.nature.com/articles/s41567-025-02889-7},
	doi = {10.1038/s41567-025-02889-7},
	language = {en},
	number = {7},
	urldate = {2025-12-17},
	journal = {Nature Physics},
	author = {Merboldt, Marco and Schüler, Michael and Schmitt, David and Bange, Jan Philipp and Bennecke, Wiebke and Gadge, Karun and Pierz, Klaus and Schumacher, Hans Werner and Momeni, Davood and Steil, Daniel and Manmana, Salvatore R. and Sentef, Michael A. and Reutzel, Marcel and Mathias, Stefan},
	month = jul,
	year = {2025},
	pages = {1093--1099},
}

@article{wang_observation_2011,
	title = {Observation of a {Warped} {Helical} {Spin} {Texture} in \$\{{\textbackslash}mathrm\{{Bi}\}\}\_\{2\}\{{\textbackslash}mathrm\{{Se}\}\}\_\{3\}\$ from {Circular} {Dichroism} {Angle}-{Resolved} {Photoemission} {Spectroscopy}},
	volume = {107},
	url = {https://link.aps.org/doi/10.1103/PhysRevLett.107.207602},
	doi = {10.1103/PhysRevLett.107.207602},
	number = {20},
	urldate = {2025-12-17},
	journal = {Physical Review Letters},
	author = {Wang, Y. H. and Hsieh, D. and Pilon, D. and Fu, L. and Gardner, D. R. and Lee, Y. S. and Gedik, N.},
	month = nov,
	year = {2011},
	pages = {207602},
}

@article{unzelmann_momentum-space_2021,
	title = {Momentum-space signatures of {Berry} flux monopoles in the {Weyl} semimetal {TaAs}},
	volume = {12},
	copyright = {2021 The Author(s)},
	issn = {2041-1723},
	url = {https://www.nature.com/articles/s41467-021-23727-3},
	doi = {10.1038/s41467-021-23727-3},
	language = {en},
	number = {1},
	urldate = {2025-12-17},
	journal = {Nature Communications},
	author = {Ünzelmann, M. and Bentmann, H. and Figgemeier, T. and Eck, P. and Neu, J. N. and Geldiyev, B. and Diekmann, F. and Rohlf, S. and Buck, J. and Hoesch, M. and Kalläne, M. and Rossnagel, K. and Thomale, R. and Siegrist, T. and Sangiovanni, G. and Sante, D. Di and Reinert, F.},
	month = jun,
	year = {2021},
	pages = {3650},
}

@article{schuler_local_2020,
	title = {Local {Berry} curvature signatures in dichroic angle-resolved photoelectron spectroscopy from two-dimensional materials},
	volume = {6},
	url = {https://www.science.org/doi/abs/10.1126/sciadv.aay2730},
	doi = {10.1126/sciadv.aay2730},
	number = {9},
	journal = {Science Advances},
	author = {Schüler, Michael and Giovannini, Umberto De and Hübener, Hannes and Rubio, Angel and Sentef, Michael A. and Werner, Philipp},
	year = {2020},
	pages = {eaay2730},
}

@article{ryoo_lippmann-schwinger_2025,
	title = {Lippmann-{Schwinger} {Approach} for {Accurate} {Photoelectron} {Wave} {Functions} and {Angle}-{Resolved} {Photoemission} {Spectra} from {First} {Principles}},
	volume = {135},
	url = {https://link.aps.org/doi/10.1103/gwmm-6l57},
	doi = {10.1103/gwmm-6l57},
	number = {5},
	urldate = {2025-12-17},
	journal = {Physical Review Letters},
	author = {Ryoo, Ji Hoon and Park, Cheol-Hwan},
	month = aug,
	year = {2025},
	pages = {056403},
}

@article{figgemeier_imaging_2025,
	title = {Imaging {Orbital} {Vortex} {Lines} in {Three}-{Dimensional} {Momentum} {Space}},
	volume = {15},
	url = {https://link.aps.org/doi/10.1103/PhysRevX.15.011032},
	doi = {10.1103/PhysRevX.15.011032},
	number = {1},
	urldate = {2025-12-17},
	journal = {Physical Review X},
	author = {Figgemeier, T. and Ünzelmann, M. and Eck, P. and Schusser, J. and Crippa, L. and Neu, J. N. and Geldiyev, B. and Kagerer, P. and Buck, J. and Kalläne, M. and Hoesch, M. and Rossnagel, K. and Siegrist, T. and Lim, L.-K. and Moessner, R. and Sangiovanni, G. and Di Sante, D. and Reinert, F. and Bentmann, H.},
	month = feb,
	year = {2025},
	pages = {011032},
}

@article{bosoni_how_2024,
	title = {How to verify the precision of density-functional-theory implementations via reproducible and universal workflows},
	volume = {6},
	copyright = {2023 Springer Nature Limited},
	issn = {2522-5820},
	url = {https://www.nature.com/articles/s42254-023-00655-3},
	doi = {10.1038/s42254-023-00655-3},
	language = {en},
	number = {1},
	urldate = {2025-12-17},
	journal = {Nature Reviews Physics},
	author = {Bosoni, Emanuele and Beal, Louis and Bercx, Marnik and Blaha, Peter and Blügel, Stefan and Bröder, Jens and Callsen, Martin and Cottenier, Stefaan and Degomme, Augustin and Dikan, Vladimir and Eimre, Kristjan and Flage-Larsen, Espen and Fornari, Marco and Garcia, Alberto and Genovese, Luigi and Giantomassi, Matteo and Huber, Sebastiaan P. and Janssen, Henning and Kastlunger, Georg and Krack, Matthias and Kresse, Georg and Kühne, Thomas D. and Lejaeghere, Kurt and Madsen, Georg K. H. and Marsman, Martijn and Marzari, Nicola and Michalicek, Gregor and Mirhosseini, Hossein and Müller, Tiziano M. A. and Petretto, Guido and Pickard, Chris J. and Poncé, Samuel and Rignanese, Gian-Marco and Rubel, Oleg and Ruh, Thomas and Sluydts, Michael and Vanpoucke, Danny E. P. and Vijay, Sudarshan and Wolloch, Michael and Wortmann, Daniel and Yakutovich, Aliaksandr V. and Yu, Jusong and Zadoks, Austin and Zhu, Bonan and Pizzi, Giovanni},
	month = jan,
	year = {2024},
	pages = {45--58},
}

@article{strocov_high-energy_2023,
	title = {High-energy photoemission final states beyond the free-electron approximation},
	volume = {14},
	issn = {2041-1723},
	url = {https://www.nature.com/articles/s41467-023-40432-5},
	doi = {10.1038/s41467-023-40432-5},
	number = {1},
	urldate = {2025-12-17},
	journal = {Nature Communications},
	author = {Strocov, V. N. and Lev, L. L. and Alarab, F. and Constantinou, P. and Wang, X. and Schmitt, T. and Stock, T. J. Z. and Nicolaï, L. and Očenášek, J. and Minár, J.},
	year = {2023},
	pages = {4827},
}

@article{gierz_graphene_2012,
	title = {Graphene {Sublattice} {Symmetry} and {Isospin} {Determined} by {Circular} {Dichroism} in {Angle}-{Resolved} {Photoemission} {Spectroscopy}},
	volume = {12},
	issn = {1530-6984},
	url = {https://doi.org/10.1021/nl300512q},
	doi = {10.1021/nl300512q},
	number = {8},
	journal = {Nano Letters},
	author = {Gierz, Isabella and Lindroos, Matti and Höchst, Hartmut and Ast, Christian R. and Kern, Klaus},
	month = aug,
	year = {2012},
	pages = {3900--3904},
}

@article{heider_geometry-induced_2023,
	title = {Geometry-{Induced} {Spin} {Filtering} in {Photoemission} {Maps} from \$\{{\textbackslash}mathrm\{{WTe}\}\}\_\{2\}\$ {Surface} {States}},
	volume = {130},
	url = {https://link.aps.org/doi/10.1103/PhysRevLett.130.146401},
	doi = {10.1103/PhysRevLett.130.146401},
	number = {14},
	urldate = {2025-12-17},
	journal = {Physical Review Letters},
	author = {Heider, Tristan and Bihlmayer, Gustav and Schusser, Jakub and Reinert, Friedrich and Minár, Jan and Blügel, Stefan and Schneider, Claus M. and Plucinski, Lukasz},
	month = apr,
	year = {2023},
	pages = {146401},
}

@article{cho_experimental_2018,
	title = {Experimental {Observation} of {Hidden} {Berry} {Curvature} in {Inversion}-{Symmetric} {Bulk} {2H}-{WSe}₂},
	volume = {121},
	url = {https://link.aps.org/doi/10.1103/PhysRevLett.121.186401},
	doi = {10.1103/PhysRevLett.121.186401},
	number = {18},
	journal = {Phys. Rev. Lett.},
	author = {Cho, Soohyun and Park, Jin-Hong and Hong, Jisook and Jung, Jongkeun and Kim, Beom Seo and Han, Garam and Kyung, Wonshik and Kim, Yeongkwan and Mo, S.-K. and Denlinger, J. D. and Shim, Ji Hoon and Han, Jung Hoon and Kim, Changyoung and Park, Seung Ryong},
	month = oct,
	year = {2018},
	pages = {186401},
}

@article{onida_electronic_2002,
	title = {Electronic excitations: density-functional versus many-body {Green}'s-function approaches},
	volume = {74},
	shorttitle = {Electronic excitations},
	url = {https://link.aps.org/doi/10.1103/RevModPhys.74.601},
	doi = {10.1103/RevModPhys.74.601},
	number = {2},
	urldate = {2025-12-17},
	journal = {Reviews of Modern Physics},
	author = {Onida, Giovanni and Reining, Lucia and Rubio, Angel},
	month = jun,
	year = {2002},
	pages = {601--659},
}

@article{borghetti_effect_2012,
	title = {Effect of surface reconstruction on the photoemission cross-section of the {Au}(111) surface state},
	volume = {24},
	issn = {0953-8984},
	url = {https://doi.org/10.1088/0953-8984/24/39/395006},
	doi = {10.1088/0953-8984/24/39/395006},
	language = {en},
	number = {39},
	urldate = {2025-12-17},
	journal = {Journal of Physics: Condensed Matter},
	author = {Borghetti, Patrizia and Lobo-Checa, Jorge and Goiri, Elizabeth and Mugarza, Aitor and Schiller, Frederik and Enrique Ortega, J and Krasovskii, Eugene E},
	month = sep,
	year = {2012},
	pages = {395006},
}

@article{yen_controllable_2024,
	title = {Controllable orbital angular momentum monopoles in chiral topological semimetals},
	volume = {20},
	copyright = {2024 The Author(s)},
	issn = {1745-2481},
	url = {https://www.nature.com/articles/s41567-024-02655-1},
	doi = {10.1038/s41567-024-02655-1},
	language = {en},
	number = {12},
	urldate = {2025-12-17},
	journal = {Nature Physics},
	author = {Yen, Yun and Krieger, Jonas A. and Yao, Mengyu and Robredo, Iñigo and Manna, Kaustuv and Yang, Qun and McFarlane, Emily C. and Shekhar, Chandra and Borrmann, Horst and Stolz, Samuel and Widmer, Roland and Gröning, Oliver and Strocov, Vladimir N. and Parkin, Stuart S. P. and Felser, Claudia and Vergniory, Maia G. and Schüler, Michael and Schröter, Niels B. M.},
	month = dec,
	year = {2024},
	pages = {1912--1918},
}

@article{nozaki_computational_2024,
	title = {Computational method for angle-resolved photoemission spectra from repeated-slab band structure calculations},
	volume = {110},
	url = {https://link.aps.org/doi/10.1103/PhysRevB.110.195406},
	doi = {10.1103/PhysRevB.110.195406},
	number = {19},
	urldate = {2025-12-17},
	journal = {Physical Review B},
	author = {Nozaki, Misa and Krüger, Peter},
	month = nov,
	year = {2024},
	pages = {195406},
}

@article{wang_circular_2013,
	title = {Circular dichroism in angle-resolved photoemission spectroscopy of topological insulators},
	volume = {7},
	copyright = {Copyright © 2013 WILEY-VCH Verlag GmbH \& Co. KGaA, Weinheim},
	issn = {1862-6270},
	url = {https://onlinelibrary.wiley.com/doi/abs/10.1002/pssr.201206458},
	doi = {10.1002/pssr.201206458},
	language = {en},
	number = {1-2},
	urldate = {2025-12-17},
	journal = {physica status solidi (RRL) – Rapid Research Letters},
	author = {Wang, Yihua and Gedik, Nuh},
	year = {2013},
	pages = {64--71},
}

@article{brinkman_chirality-driven_2024,
	title = {Chirality-{Driven} {Orbital} {Angular} {Momentum} and {Circular} {Dichroism} in {CoSi}},
	volume = {132},
	url = {https://link.aps.org/doi/10.1103/PhysRevLett.132.196402},
	doi = {10.1103/PhysRevLett.132.196402},
	number = {19},
	urldate = {2025-12-17},
	journal = {Physical Review Letters},
	author = {Brinkman, Stefanie Suzanne and Tan, Xin Liang and Brekke, Bj{\o}rnulf and Mathisen, Anders Christian and Finnseth, {\O}yvind and Schenk, Richard Justin and Hagiwara, Kenta and Huang, Meng-Jie and Buck, Jens and Kall{\"a}ne, Matthias and Hoesch, Moritz and Rossnagel, Kai and Ou Yang, Kui-Hon and Lin, Minn-Tsong and Shu, Guo-Jiun and Chen, Ying-Jiun and Tusche, Christian and Bentmann, Hendrik},
	month = may,
	year = {2024},
	pages = {196402},
}

@article{krasovskii_calculation_1999,
	title = {Calculation of the wave functions for semi-infinite crystals with linear methods of band theory},
	volume = {59},
	url = {https://link.aps.org/doi/10.1103/PhysRevB.59.R15609},
	doi = {10.1103/PhysRevB.59.R15609},
	number = {24},
	journal = {Phys. Rev. B},
	author = {Krasovskii, E. E. and Schattke, W.},
	month = jun,
	year = {1999},
	pages = {R15609--R15612},
}

@article{erhardt_bias-free_2024,
	title = {Bias-{Free} {Access} to {Orbital} {Angular} {Momentum} in {Two}-{Dimensional} {Quantum} {Materials}},
	volume = {132},
	url = {https://link.aps.org/doi/10.1103/PhysRevLett.132.196401},
	doi = {10.1103/PhysRevLett.132.196401},
	number = {19},
	journal = {Phys. Rev. Lett.},
	author = {Erhardt, Jonas and Schmitt, Cedric and Eck, Philipp and Schmitt, Matthias and Keßler, Philipp and Lee, Kyungchan and Kim, Timur and Cacho, Cephise and Cojocariu, Iulia and Baranowski, Daniel and Feyer, Vitaliy and Veyrat, Louis and Sangiovanni, Giorgio and Claessen, Ralph and Moser, Simon},
	month = may,
	year = {2024},
	pages = {196401},
}

@article{takahashi_berry_2015,
	title = {Berry curvature and orbital angular momentum of electrons in angle-resolved photoemission spectroscopy},
	volume = {91},
	url = {https://link.aps.org/doi/10.1103/PhysRevB.91.245133},
	doi = {10.1103/PhysRevB.91.245133},
	number = {24},
	journal = {Phys. Rev. B},
	author = {Takahashi, Ryuji and Nagaosa, Naoto},
	month = jun,
	year = {2015},
	pages = {245133},
}

@article{krasovskii_band_2007,
	title = {Band mapping in the one-step photoemission theory: {Multi}-{Bloch}-wave structure of final states and interference effects},
	volume = {75},
	url = {https://link.aps.org/doi/10.1103/PhysRevB.75.045432},
	doi = {10.1103/PhysRevB.75.045432},
	number = {4},
	journal = {Phys. Rev. B},
	author = {Krasovskii, E. E. and Strocov, V. N. and Barrett, N. and Berger, H. and Schattke, W. and Claessen, R.},
	month = jan,
	year = {2007},
	pages = {045432},
}

@article{schusser_assessing_2022,
	title = {Assessing {Nontrivial} {Topology} in {Weyl} {Semimetals} by {Dichroic} {Photoemission}},
	volume = {129},
	url = {https://link.aps.org/doi/10.1103/PhysRevLett.129.246404},
	doi = {10.1103/PhysRevLett.129.246404},
	number = {24},
	urldate = {2025-12-17},
	journal = {Physical Review Letters},
	author = {Schusser, J. and Bentmann, H. and Ünzelmann, M. and Figgemeier, T. and Min, C.-H. and Moser, S. and Neu, J. N. and Siegrist, T. and Reinert, F.},
	month = dec,
	year = {2022},
	pages = {246404},
}

@article{sobota_angle-resolved_2021,
	title = {Angle-resolved photoemission studies of quantum materials},
	volume = {93},
	url = {https://link.aps.org/doi/10.1103/RevModPhys.93.025006},
	doi = {10.1103/RevModPhys.93.025006},
	number = {2},
	journal = {Rev. Mod. Phys.},
	author = {Sobota, Jonathan A. and He, Yu and Shen, Zhi-Xun},
	month = may,
	year = {2021},
	pages = {025006},
}

@article{krasovskii_ab_2021,
	title = {Ab {Initio} {Theory} of {Photoemission} from {Graphene}},
	volume = {11},
	copyright = {http://creativecommons.org/licenses/by/3.0/},
	issn = {2079-4991},
	url = {https://www.mdpi.com/2079-4991/11/5/1212},
	doi = {10.3390/nano11051212},
	language = {en},
	number = {5},
	urldate = {2025-12-17},
	journal = {Nanomaterials},
	author = {Krasovskii, Eugene},
	month = may,
	year = {2021},
}

\end{document}